# Experimental demonstration of stimulated polarization wave in a chain of nuclear spins

Jae-Seung Lee, Travis Adams, and A. K. Khitrin

Department of Chemistry, Kent State University, Kent, Ohio 44242-0001

**Abstract**

A stimulated wave of polarization, which implements a simple mechanism of quantum amplification, is experimentally demonstrated in a chain of four J-coupled nuclear spins, irradiated by a weak radio-frequency transverse field. The "quantum domino" dynamics, a wave of flipped spins triggered by a flip of the first spin, has been observed in fully $^{13}$C-labeled sodium butyrate.

In quantum dynamics, governed by linear equations of motion, amplification can be realized by using entangled quantum states of a system. In this case, a local perturbation, affecting a small part of the system, changes a wave function of the entire system in a coherent way. Subsequently, this change may be converted into changes of "macroscopic" observables [1-3]. For an $N$-quibit system, one of simple logical schemes of this process is a chain of unitary controlled-NOT operations [3]

$$U = CNOT_{N-1,N} CNOT_{N-2,N-1} \ldots CNOT_{2,3} CNOT_{1,2}, \qquad (1)$$

where $CNOT_{m,n}$ flips the state of the $n$-th qubit if the $m$-th qubit is in the state $|1\rangle$ and does not do anything if the state is $|0\rangle$. The unitary transformation (1) converts the initial state

$$|\psi_{in}\rangle = (a|0\rangle_1 + b|1\rangle_1)|0\rangle_2|0\rangle_3 \ldots |0\rangle_{N-1}|0\rangle_N \qquad (2)$$

into the final state

$$|\psi_{out}\rangle = U|\psi_{in}\rangle = a|0\rangle_1|0\rangle_2 \ldots |0\rangle_{N-1}|0\rangle_N + b|1\rangle_1|1\rangle_2 \ldots |1\rangle_{N-1}|1\rangle_N \qquad (3)$$

where the quantum state of the first qubit is "expanded", and the state of polarization of this single qubit is transferred to the state of the total polarization of the entire cluster.

If viewed as consecutive steps in time, one gate after another, the chain (1) describes a "quantum domino" dynamics, when a wave of flipped qubits is triggered by a flip of the first qubit. It is also possible to create schemes where the number of flipped spins grows faster than linear with the number of logic steps [4]. A physical model with continuous dynamics, similar to that suggested by the chain (1), is an Ising chain with nearest-neighbor interactions, irradiated by a weak resonant transverse field [5]. The Hamiltonian of this model is

$$H = \frac{\omega_0}{2}\sum_{i=1}^{N}\sigma_i^z + \omega_1\sum_{i=1}^{N}\sigma_i^x \cos\omega_0 t + \frac{J}{4}\sum_{i=1}^{N-1}\sigma_i^z\sigma_{i+1}^z, \qquad (4)$$

where $\omega_0$ is the energy difference ($\hbar = 1$) between the excited and ground states of an isolated spin (qubit), $J$ is the interaction constant, $\omega_1 \ll J \ll \omega_0$ is the amplitude of irradiation, $\sigma^z$ and $\sigma^x$ are the Pauli operators. Qualitatively, the principle of operation of the model with the Hamiltonian (4) can be explained as follows. The weak resonant field with amplitude $\omega_1$ can flip a spin only if its two neighbors are in different states and the shifts of the resonance frequency, caused by interaction with these neighbors, are compensated. Therefore, if the initial state is all spins up, nothing happens. If the first spin is flipped, its neighbor becomes resonant and flips, then the next neighbor, and so on, generating the polarization wave. Of course, real multi-spin dynamics is more complex and all spins move simultaneously. In the limit $\omega_1 \to 0$ an analytical solution [5] is available for this model. Its results support the qualitative picture described above.

It is difficult to find a real physical system with the Hamiltonian (4). Simulations for short spin chains with more realistic Hamiltonians [6] showed that, even in the absence of relaxation and decoherence, the possibility to launch a strong polarization wave and reach efficient amplification critically depends on parameters of the spin Hamiltonian. Therefore, experimental demonstration of this phenomenon would be encouraging.

In liquid-state NMR, it is possible to find spin systems with Hamiltonians resembling (4). Isotropic J-couplings between the nearest spins are much stronger than between the remote spins. Therefore, a linear chain of nuclear spins may be almost a chain with the nearest-neighbor interactions. Truncation of the isotropic J-coupling to a ZZ-term can result from large difference between the chemical shifts of the neighbor spins. An unavoidable complication is a need for multi-frequency irradiation to irradiate each spin at its own resonance frequency. For the present experiment, we have chosen a chain of four $^{13}$C nuclear spins of fully $^{13}$C-labeled sodium butyrate. The major differences between its spin Hamiltonian (under proton decoupling) and the Hamiltonian (4) are the following. All spins have different resonance frequencies. J-coupling constants between the nearest neighbors are not all equal. There are small couplings between the next- and next-next-nearest neighbors.

The thermal equilibrium $^{13}$C spectrum is shown in Fig.1, line (a). Interaction with the nearest neighbors splits the spectra of spins 1 and 4 into doublets, and that of spin 3 into triplet. Four peaks of spin 2 result from different couplings to spins 1 and 3. Fine structure of the peaks comes from interactions beyond the nearest neighbors.

The first step of our experiment is the initialization of the system in the state with all spins up. This pseudopure state [7,8] has been prepared by using a partial saturation [9]. The saturation has been performed with a seven-frequency pulse, which irradiated all allowed single-quantum transitions, except for the transitions with frequencies close to the transitions from the ground state (all spins up) to the four states with one flipped spin. As a result, there were no transitions from the ground state, and its population has been "trapped", while the populations of the other 15 states have been equalized. The saturating pulse was 75 ms long and had an amplitude ($\gamma B_1/2\pi$) of 19 Hz per harmonic. A linear-response spectrum of the pseudopure ground state is presented in Fig. 1(b). It contains four peaks, one per spin, corresponding to transitions to the states with one flipped spin. The quality of the state is supported by the fact that the multiplet structure of the peaks disappeared. It is interesting that the peaks close to those in Fig. 1(b) have been eliminated even though these transitions were not saturated directly.

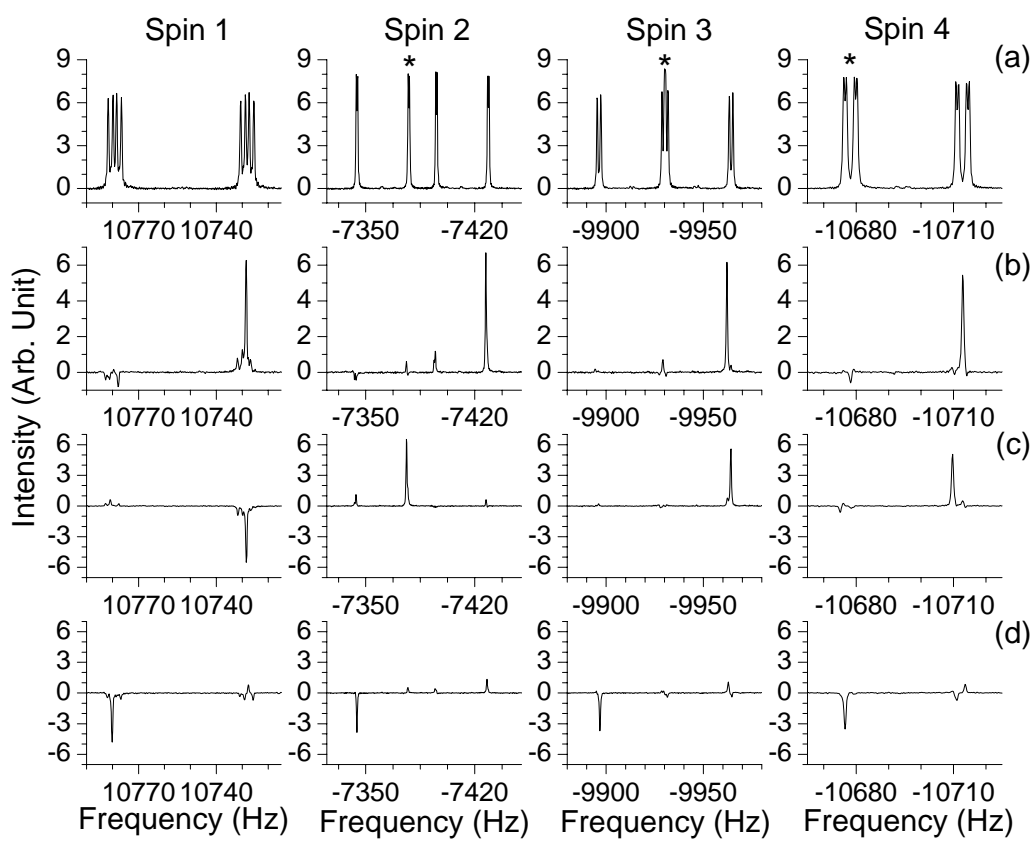

Fig. 1. Line (a): thermal equilibrium $^{13}$C NMR spectra for the spins 1 to 4; (b) pseudopure state with all spins up; (c) the spectra after the first spin has been flipped; (d) the spectra after the evolution time $\omega_1 t = 5.2$.

The result of flipping spin 1 with selective Gaussian pulse is shown in Fig. 1(c). It is a linear-response spectrum for the state with spin 1 down and spins 2-4 up. One can see that, compared to Fig. 1(b), the peak for spin 2 "shifted" to the left. This shift can be viewed as resulting from a change of the sign of the exchange field created by spin 1 when spin 1 flipped. One can also notice a small shift for spin 4. It results from the fact that the next-strongest interaction for spin 4, after its interaction with spin 3, is the interaction with spin 1, instead of spin 2.

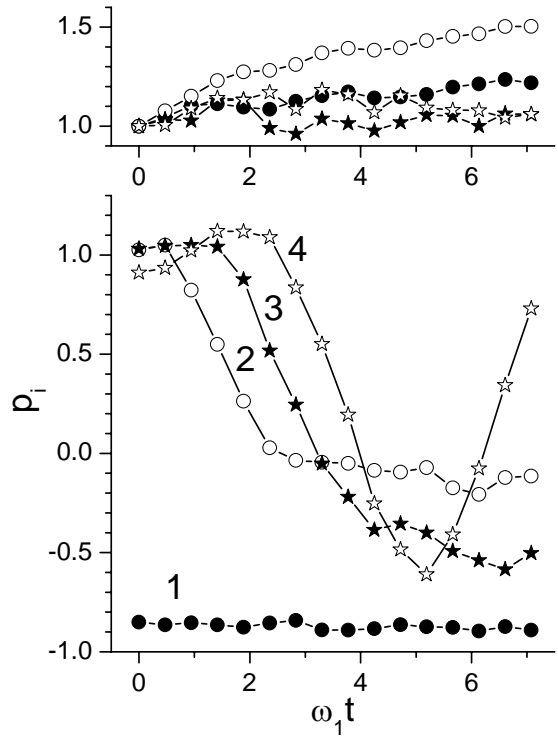

Fig. 2. Polarizations of individual spins after the first spin has been flipped (bottom), and without flipping the first spin (top).

The evolution has been driven by a three-frequency pulse with incremented length 0 to 150 ms. The three frequencies of irradiation marked by asterisks in Fig. 1(a) are the resonance frequencies for the spins 2-4 in the states where the left neighbor spin is down and the right neighbor spin (for spins 2 and 3) is up. The amplitude of the evolution pulse was 7.5 Hz for each harmonic. Evolution of polarizations of individual spins is shown in the bottom of Fig. 2. Spin polarizations have been measured by integrating the spectra in the entire spectral range for each of the spins and comparison with the pseudopure ground state in Fig. 1(b). One can clearly see the wave of flipped spins in Fig. 2, when first the spin 2, then spin 3, and finally spin 4 flip. A linear-response spectrum after the evolution time $\omega_1 t = 5.2$ in Fig. 1(d) shows that in this state all spins are down. The result of the same experiment without flipping spin 1 is shown in the top of Fig. 2. Ideally, we would expect no changes in this case. Slow growth of polarizations is due to spin-lattice relaxation and the nuclear Overhauser effect, resulting from the proton decoupling.

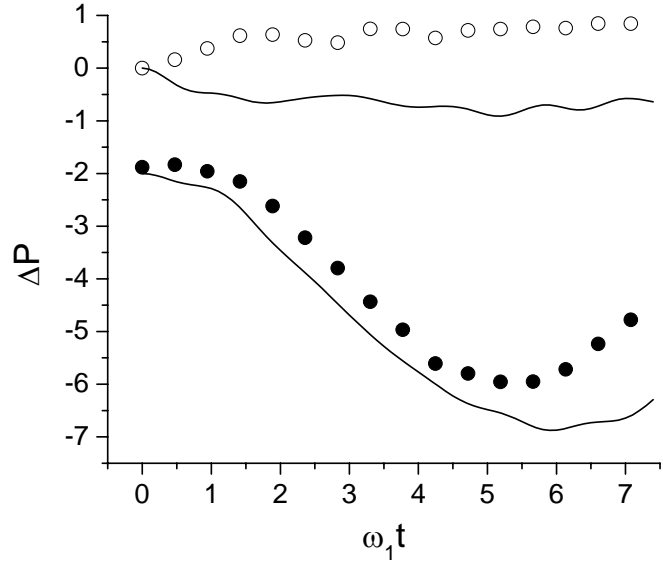

Fig. 3. Evolution of the total polarization, triggered by a flip of the first spin (filled circles), and without flipping the first spin (open circles); the lines are the results of numerical simulation.

Evolution of the total polarization of the four-spin system, with and without flip of the first spin, is shown in Fig. 3. The lines are the results of computer simulation. Deviations between the experimental and simulated dynamics are due to the relaxation and the Overhauser effect. An efficiency of the amplification dynamics can be described by a coefficient of amplification [5] defined as the ratio of the maximum change of polarization in a dynamics, triggered by a flip of a spin, to a direct change in polarization by a flip of a single spin. In our experimental implementation the measured coefficient of amplification is about 3.

In conclusion, a stimulated wave of polarization, triggered by a flip of the end spin, has been experimentally observed in a linear chain of nuclear spins. This "quantum domino" dynamics is an explicit realization of a mechanism of quantum amplification.

The work was supported by NSF and Ohio Board of Regents.